\documentclass[aps,showpacs,twocolumn,floats,epsfig,pdflatex]{revtex4}
\usepackage{epsfig}
\usepackage{amsmath}
\usepackage{amsfonts}
\usepackage{graphicx}
\usepackage{amssymb}
\usepackage{amsbsy}
\usepackage{subfigure}

\begin{document}
\title {Effect of spin-orbit interaction on the critical temperature of an ideal Bose gas }
\author{Arunesh Roy, Sayak Ray and S. Sinha}

\affiliation{Indian Institute of Science Education and
Research-Kolkata, Mohanpur, Nadia 741252, India.}

\date{\today}
\begin{abstract}
We consider Bose-Einstein condensation of an ideal bose gas with an equal mixture of `Rashba' and `Dresselhaus' spin-orbit interactions and study its effect on the critical temperature.
 In uniform bose gas a `cusp' and a sharp drop in the critical temperature occurs due to the change in the density of states at a critical Raman coupling where the degeneracy of the ground states is lifted.  Relative drop in the critical temperature depends on the diluteness of the gas as well as on the spin-orbit coupling strength. In the presence of a harmonic trap, the cusp in the critical temperature smoothened out and a minimum appears.
Both the drop in the critical temperature and lifting of `quasi-degeneracy' of the ground states exhibit crossover phenomena which is controlled by the trap frequency. By considering a 'Dicke' like model we extend our calculation to bosons with large spin and observe a similar minimum in the critical temperature near the critical Raman frequency, which becomes deeper for larger spin. Finally in the limit of infinite spin, the critical temperature vanishes at the critical frequency, which is a manifestation of Dicke type quantum phase transition.
\end{abstract}


\maketitle 
\section{Introduction}

In a recent seminal experiment, the creation of synthetic `spin-orbit'(SO) interactions in an atomic condensate\cite{spielman1} has generated an impetus to study the effects of spin-orbit interactions in the many-body physics of ultracold quantum gases\cite{spielman2}.  The SO interaction in electronic systems arises due to the relativistic corrections and the interaction strength is much smaller compared to the average kinetic energy of the electrons. Most common spin-orbit interactions that arises in solid-state materials are `Rashba'\cite{rashba} and `Dresselhaus' spin-orbit interactions\cite{dressel} with coupling constants which are fixed by the material properties. 
On the contrary, the artificial SO interaction generated in cold atoms can have both `Rashba' and `Dresselhaus' term with any proportion and the strength of SO couplings can be tuned by the quantum engineering techniques\cite{ohberg, dalibard}. The SO coupled cold atoms opened up  new possibilities to study various types of exotic quantum phases like `stripe phase', `half-vortex', `density wave', `hexagonal lattice phase' etc\cite{zhai1,ho,wu,machida,sinha,pu}.
Moreover, the time-reversal symmetry of the  SO interaction  is also an important ingredient for the formation of 'topological matter'\cite{kane,SHE,spielman2}. Exotic correlated quantum phases and spin textures can be realized in cold atoms with SO interaction in an optical lattice\cite{radic,cole,grass}. In recent years, a significant amount of work has been done to explore various interesting physical aspects of SO coupled ultracold atoms (for a review, see \cite{dalibard,zhai2}).

In the first experiment, SO interaction with an equal mixture of Rashba and Dresselhaus type has been created in a condensate of $^{87}Rb$ atoms\cite{spielman1}. In this case, the single particle energy dispersion shows two degenerate minima at finite momentum. In addition to the SO interaction two pseudospin states of the atom are coupled with Raman frequency $\Omega$. By increasing the Raman frequency the energy spectrum changes and above a critical value of the Raman frequency, degeneracy of the ground state vanishes and a single ground state at zero momentum appears. In general, the SO interaction leads to the degenerate ground states which makes the condensate an `unconventional' type. In case of Rashba spin-orbit coupling the infinite degeneracy of the ground states strongly modifies the single particle density of states(DOS) which in turn affects the Bose-Einstein transition\cite{rashba_tc}. In the presence of inter particle interactions, the condensate can break the symmetry and choose one of the ground states whose thermodynamic stability is an interesting aspect\cite{baym1,barnett}.

In the presence of inter particle interactions, the condensate with experimentally generated SO interaction, has three different phases: i)a homogeneous condensate with zero momentum ii) time reversal symmetry broken condensate with finite momentum and iii) superstripe phase, which is a linear superposition of equal and opposite momentum states\cite{ho,stringari1}. In this particular case, the condensate undergoes a quantum phase transition at zero temperature by tuning the SO coupling strength or Raman frequency. In a recent experiment the finite temperature phase diagram of this system has been obtained which revealed various interesting features of the above mentioned  phases\cite{ji}. The effects of SO coupling and inter particle interaction are reflected in the critical temperature and condensate fraction.
 Interaction induced shift in the critical temperature, depletion of the condensate in presence of SO coupling have also been studied theoretically\cite{zhai3}.

Motivated by the recent experiments\cite{spielman1,ji}, in this work we consider the Bose-Einstein condensation(BEC) of simple non-interacting bosons with equal mixture of `Rashba' and `Dresselhaus' SO interaction for both homogeneous and harmonically trapped system. In absence of interparticle interactions, last two phases (ii and iii, as mentioned above) do not exist and the system becomes relatively simple. Depending on the Raman frequency bosons can have macroscopic occupation at single non degenerate ground state or can be equally distributed in two degenerate ground states with equal and opposite momentum. This is a simple and straight forward extension of the text book material. However this simple calculation can capture some of the interesting features of the critical temperature which has already been observed experimentally\cite{ji} and provides a clear physical picture behind it.
We analyze the appearance of a cusp and a sharp drop of the critical temperature at a critical Raman coupling where the single particle energy spectrum changes its shape.
This phenomena is closely related to the degeneracy lifting of the ground state, since at the critical Raman coupling two degenerate ground states with equal and opposite momentum vanishes and a single ground state with zero momentum appears. The relative drop of critical temperature depends on the ratio between interparticle separation and the length scale of SO coupling. The critical temperature decreases for increasing diluteness of the gas. In the presence of a harmonic trap, the sharp `cusp' in critical temperature is smoothened out and exhibits a crossover phenomenon which is controlled by the trap frequency. We also observe a similar crossover phenomenon in degeneracy lifting of the ground states  by analyzing the ground state energy gap.
We also extend our calculation of critical temperature for harmonically trapped bosons with large spin. The drop in the critical temperature increases for increasing spin of the bosons. In the limit of infinite spin, the critical temperature vanishes at critical Raman coupling revealing a `Dicke' type quantum phase transition related to the spin ordering. 
  
This paper is organized as follows. In Sec.II, we study the single particle dispersion relation of free bosons with SO interaction and calculate the critical temperature. In presence of a harmonic trap semiclassical calculation for the critical temperature is presented in Sec.III. This is followed by an analysis to study the degeneracy lifting of the ground state of trapped bosons by calculating the energy gap above the ground state. In Sec.IV,  we calculate the critical temperature of bosons with large spin by analyzing a `Dicke' like model. Finally the results are summarized in Sec. V.   
 
\section{Critical temperature of uniform bose gas with SO interaction}
The single particle Hamiltonian of the spin-orbit coupled atoms in presence of Rabi coupling is given by,
\begin{equation}
H =  \frac{\hat{\vec{p}}^2}{2m} + \frac{\hbar k_{L}}{m}\hat{p}_{x} \sigma_{z} + \hbar \Omega \sigma_{x} + \frac{\delta}{2}\sigma_{z} + V(\vec{r}),
\label{hamil_f}
\end{equation}
where $m$ is the mass of the atoms and $\sigma_{x,y,z}$ are $2\times 2$ Pauli matrices. The first term of the Hamiltonian describes the usual kinetic energy of the atoms, and the second term represents an equal combination of Rashba and Dresselhaus spin-orbit interactions. The coupling strength of the SO interaction is determined by the tunable wavevector $k_{L}$ of the laser. The third term of the Hamiltonian describes coupling between two internal states (ground and excited states) of an atom with frequency $\Omega$. Last term gives the energy difference between two internal states due to detuning and for simplicity we can drop this term in rest of this paper. The atoms are confined by the potential $V(\vec{r})$ due to trap.

In the absence of a trapping potential, above Hamiltonian can be diagonalized by the plane wave basis states. After diagonalizing the $2\times 2$ Hamiltonian in momentum basis, two branches of the energy spectrum are given by,
\begin{equation}
E(\vec{k})_{\pm} = \frac{\hbar^2 k^2}{2m} \pm \sqrt{\left(\frac{\hbar^2 k_{x} k_{L}}{m}\right)^2 + \hbar^2 \Omega^2}.
\label{energy_f}
\end{equation}
For a plane wave state with wavevector $\vec{k}$, two component spinors corresponding to the two energy branches are given by,\\
\begin{eqnarray}
\psi_{+} = \frac{1}{\sqrt{V}}e^{i \vec{k}.\vec{r}}\left( \begin{array}{c} \cos \theta(k_{x})\\ \sin \theta(k_{x})\\ \end{array}\right),\\
\label{spinor_f1}
\psi_{-} = \frac{1}{\sqrt{V}}e^{i \vec{k}.\vec{r}}\left( \begin{array}{c} \sin \theta(k_{x})\\ -\cos \theta(k_{x})\\ \end{array}\right),
\label{spinor_f2}
\end{eqnarray}
where $V$ is the total volume, and the components of the spinor can be obtained from the relation,
\begin{eqnarray}
\sin \theta_k & = & \left\{\frac{1}{2}\left[1 - \frac{\tilde{k}_{x}}{\sqrt{\tilde{k}_{x}^2 + \eta^2}}\right]\right\}^{1/2},\\
\cos \theta_k & = & \left\{\frac{1}{2}\left[1 + \frac{\tilde{k}_{x}}{\sqrt{\tilde{k}_{x}^2 + \eta^2}}\right]\right\}^{1/2}.
\end{eqnarray}
Here we introduce a dimensionless momentum $\vec{\tilde{k}} = \vec{k}/k_{L}$, and dimensionless coupling constant $\eta = m \Omega/\hbar k_{L}^2$.
\begin{figure}[b]
\centering
\includegraphics[width=8cm]{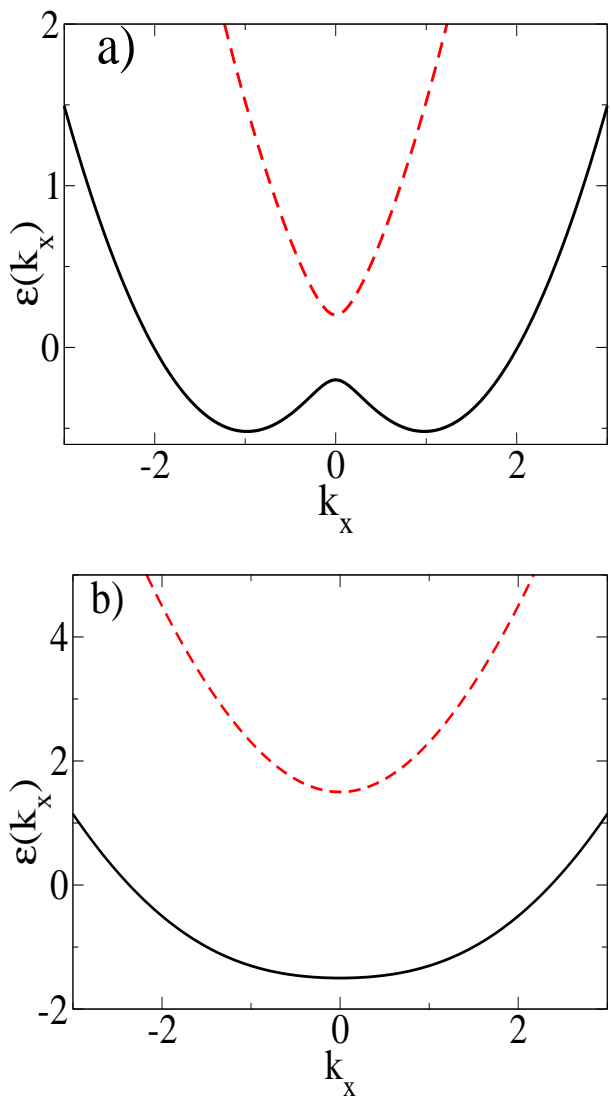}
\caption{Single particle energy (in units of $\frac{\hbar^2 k_{L}^2}{m}$)-momentum (in units of $k_{L}$) dispersion relation for a) $\eta = 0.2$ and b) $\eta = 1.5$. Solid line represents the lower branch of energy with $s=-1$ and dashed line denotes the upper branch with $s=1$.}
\label{fig1}
\end{figure}

For $\eta < 1$, the lower energy branch $\epsilon_{-}(\vec{k})$ has two degenerate minima at $\tilde{k}_{x} = \pm \sqrt{1 - \eta^2}$ as shown in Fig.1a. 
The Hamiltonian in Eq.(\ref{hamil_f}) is not invariant under the time-reversal symmetry operator due to the presence of the Raman coupling, but it is invariant under the operator $\sigma_{z} {\cal T}$, where ${\cal T} = i \sigma_{y} {\cal C}$ is the time reversal operator and ${\cal C}$ is complex conjugation operator. The degenerate ground states are related by the symmetry operator.
This degeneracy in the ground state vanishes when $\eta \ge 1$ (see Fig.1b), and there is only one ground state at $\vec{k} = 0$. This transition can be understood in a very simple way by considering spin as a classical vector in x-z plane with magnitude $S$, and making an angle $\phi$ with the z-axis. In this quasi-classical representation, the single-particle energy is given by,
\begin{equation}
\epsilon(\vec{k},\phi) = \frac{\hbar^2 k_{L}^2}{m}\left[\frac{\tilde{k}^2}{2} + \tilde{k} S \cos \phi + \eta S \sin \phi\right].
\end{equation}
From this dispersion relation we see that the ground state has momentum 
$\tilde{k} = -S\cos \phi$ and energy $\epsilon_{min} = -\frac{S^2}{2} \cos^2 \phi + S\eta \sin \phi$, which depends on the orientation of the spin. A straight forward minimization gives $\sin\phi = -\eta/S$, and the ground state is doubly degenerate at momenta $\pm \sqrt{1 - (\frac{\eta}{S})^2}$ for $\eta < S$. In this case the spin of the atom has non-vanishing component along the z-axis. For $\eta \ge S$, the spins are aligned along the x-axis ($\cos \phi =0$) and a single ground state appears at $k_{min}=0$. From this simple analysis it is clear that there is also a transition in the spin orientation of the atoms as the single particle dispersion relation changes. 

Once the single particle energies are known, thermodynamic quantities of non interacting bosons can be calculated at a finite temperature $T$. Here we follow the textbook prescription for bosons in grand canonical ensemble.
For gas of bosons the total density $\rho$ can be decomposed in two parts,
$\rho = \rho_{c} + \rho_{nc}$, where $\rho_{c}$ is the density of particles macroscopically occupying the ground state (or ground states) and $\rho_{nc}$ denotes the density of non-condensate particles occupying the excited states. For BEC the condensate density $\rho_{c}$ becomes nonzero below the critical temperature $T_{c}$. The density of non-condensate particles is given by,
\begin{eqnarray}
\rho_{nc} & = & \sum_{\vec{k} \neq \vec{k}_{min},s} \frac{1}{ze^{\beta E(\vec{k},s)} -1}\\
& = & \sum_{s}\int_{E_{min}} g_{s}(E) \frac{1}{ze^{\beta E} -1} dE
\label{denssi_nc}
\end{eqnarray}
where $z=e^{\beta \mu}$ is known as fugacity of the gas with chemical potential $\mu$, and s represents the (pseudo)spin index.  At the critical temperature the condensate density $\rho_{c}$ vanishes and the chemical potential $\mu$ approaches to the ground state energy $E_{min}$ from below. The critical temperature $T_{c}$ for BEC can be calculated from the relation,
\begin{equation}
\rho = \sum_{s}\int_{0}^{\infty} g_{s}(\epsilon) \frac{1}{e^{\beta_{c} \epsilon} -1} d\epsilon,
\label{crit_t}
\end{equation}
where energy is measured with respect to the ground state $\epsilon = E- E_{min}$, and $\beta_{c}= 1/k_{B}T_{c}$. The convergence of the above integral is required for BEC to occur at a finite non zero critical temperature. 
The density of states of bosons in three dimensions is $g(\epsilon) = \frac{2}{\sqrt{\pi}}(\frac{m}{2 \pi \hbar^2})^{3/2} \sqrt{\epsilon}$, and the critical temperature of homogeneous Bose gas is $T^{0}_{c} = \frac{2 \pi \hbar^2 n^{2/3}}{m \zeta(3/2)^{2/3}}$.

The density of states of the atoms with SO coupling can be calculated analytically from the relation,
\begin{equation}
g(E) = \frac{m V}{(2 \pi \hbar)^2}\sum_{s}\int dk_{x} \theta(E - E_{s}(k_{x})),
\label{ds_3d}
\end{equation}
where $E_{s}(k_{x})$ is the dispersion along the x-axis. 
For $\eta < 1$, using Eq.(\ref{ds_3d}) and Eq.(\ref{energy_f}) we obtain,
\begin{eqnarray}
& & g(\epsilon)  =   \frac{m V K_{L}}{\sqrt{2}(\pi \hbar)^2}\left[k_{+} -k_{-}\right],~ \text{for}~ 0<\tilde{\epsilon}\leq \frac{1}{2}(1 -\eta)^2,
\nonumber\\
& & =   \frac{m V k_{L}}{\sqrt{2}(\pi \hbar)^2} k_{+}, ~\text{for}~ \frac{1}{2}(1 -\eta)^2<\tilde{\epsilon}\leq \frac{1}{2}(1 +\eta)^2,\nonumber\\
& & = \frac{m V K_{L}}{\sqrt{2}(\pi \hbar)^2}\left[k_{+} + k_{-}\right],~\text{for}~ \frac{1}{2}(1 +\eta)^2 <\tilde{\epsilon}
\label{dosl}
\end{eqnarray}
where, $k_{\pm} = \sqrt{\tilde{\epsilon} +\frac{1}{2}(1 - \eta^2) \pm \sqrt{2\tilde{\epsilon}}}$, $\tilde{\epsilon} = \epsilon/\epsilon_{0}$, and $\epsilon_{0} = \frac{\hbar^2 k_{L}^2}{m}$.
For $\eta=0$, the DOS becomes twice the DOS of free particle, which is equivalent to that of a (pseudo)spin $1/2$ particle.  This is because the energy dispersion of the lower branch becomes two parabola at $k_{min} = \pm\sqrt{1- \eta^2}$.  For $\eta =0$, we obtain $T_{c} = T^{0}_{c}/2^{1/3}$.  For $\epsilon \ll  \frac{1}{2}(1 -\eta)^2$, the DOS can be approximated as,
\begin{equation}
g(\epsilon) =  \frac{m V k_{L}}{(\pi \hbar)^2}\frac{\sqrt{2\tilde{\epsilon}}}{\sqrt{1 - \eta^2}}.
\label{dosl1}
\end{equation}
This is equivalent to approximating the lower branch of dispersion along x by that of a free particle with an effective mass, $E_{-}(k_{x}) \approx \frac{1}{2 m^{*}}(k_{x} - k_{min})^2$, where $m^{*}/m = \frac{1}{1 - \eta^2}$.

Similarly for $\eta > 1$, the DOS is given by,
\begin{eqnarray}
g(\epsilon) & = &  \frac{m V K_{L}}{\sqrt{2}(\pi \hbar)^2} \sqrt{k_{1+}}~\text{for}~ 0 < \tilde{\epsilon} < 2 \eta\\
& = & \frac{m V K_{L}}{\sqrt{2}(\pi \hbar)^2}\left[\sqrt{k_{1+}} + \sqrt{k_{1-}}\right] ~\text{for}~ \tilde{\epsilon} > 2 \eta,
\label{dosg}
\end{eqnarray}
where, $k_{1\pm} = \tilde{\epsilon} + 1 - \eta + \sqrt{2 \tilde{\epsilon} + (\eta -1)^2}$.
In this case the double minima in the dispersion of lower branch vanishes and it can approximated by that of a free particle with an effective mass $m^{*}/m = \eta /(\eta - 1)$. Using this approximation which is valid for $\tilde{\epsilon} \ll \eta -1$, the DOS is given by,
\begin{equation}
g(\epsilon) \approx \frac{m V K_{L}}{\sqrt{2}(\pi \hbar)^2}\sqrt{\frac{\eta}{\eta -1}} \sqrt{\tilde{\epsilon}}.
\label{dosg1}
\end{equation}
In terms of the dimensionless density $\alpha = \rho/k_{L}^{3}$ the critical temperature can be obtained from the relation,
\begin{equation}
 \rho/k_{L}^{3} = \frac{1}{\pi^2}\int_{0}^{\infty}dx \frac{\tilde{g}(x)}{e^{\tilde{\beta}_{c}x} -1},
\label{criticalt1}
\end{equation}
where $\tilde{\beta}_{c} = \epsilon_{0}/k_{B}T_{c}$, $x= \epsilon/\epsilon_{0}$, and $ \tilde{g}(x) = \hbar^2 g(x)/(mk_{L} V)$. 
\begin{figure}
\rotatebox{0}{\includegraphics*[width=8cm]{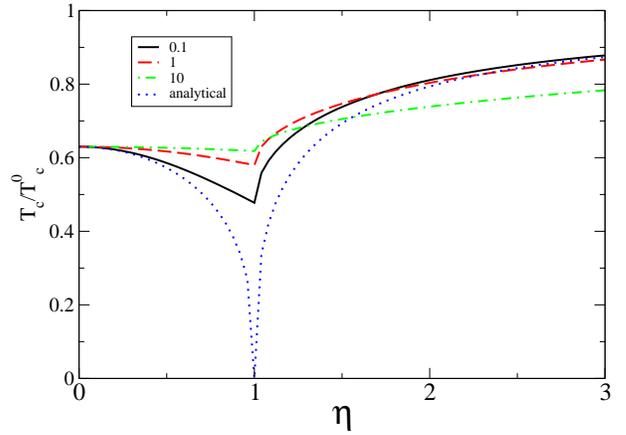}}
\caption{Variation of the scaled critical temperature $T_{c}/T^{0}_{c}$ of homogeneous bose gas with SO coupling with increasing Raman coupling $\Omega$ (in units of $\frac{\hbar^2 k_{L}^2}{m}$). Different lines represent the critical temperature for various values of dimensionless density $\alpha$ of the gas (values are indicated in the graph). Analytical expressions obtained from Eq.\ \ref{crittempl} and Eq.\ \ref{crittemp2} are represented by dotted line. }
\label{fig2}
\end{figure}

The variation of critical temperature of SO coupled bose gas with increasing Raman coupling (or $\eta$) is shown in Fig.\ \ref{fig2}. A cusp in $T_{c}$ at critical Raman coupling $\eta = 1$ appears, which is an interesting feature of SO coupling. This particular effect has been observed in recent experiment\cite{ji} and in theoretical study\cite{zhai3}. We find the drop of critical temperature is larger for bose gas with lower density as depicted in Fig.\ \ref{fig2}. Next we present analytical estimates of $T_{c}$ and its relative drop at $\eta=1$.

For strong SO interaction, only the lowest energy branch beomes important and we can obtain the approximate analytic expression for the critical temperature. For $\eta <1$, when the thermal energy is much smaller than the barrier height of the double well energy spectrum, $k_{B}T_{c} \ll \epsilon_{0} \frac{1}{2}(1 -\eta)^2$, using Eq(\ref{dosl1}) in Eq(\ref{criticalt1}), we obtain,
\begin{equation}
T_{c} =  (1 - \eta^2)^{1/3}T_{c}^{0}/2^{2/3}.
\label{crittempl}
\end{equation}
The above expression is valid for dilute gas when $\alpha \ll 1 -\eta$, and it breaks down near $\eta = 1$. 
Similarly for $\eta >1$, the critical temperature can be obtained analytically by using Eq(\ref{dosg1}) for density of states within lowest branch approximation,
\begin{equation}
T_{c} = (\frac{\eta -1}{\eta})^{1/3} T_{c}^{0}.
\label{crittemp2}
\end{equation}
This is also valid in the regime when the thermal energy is less than the gap between the energy branches $2\eta$. 
Considering only the lowest branch of dispersion, the critical temperature at $\eta =1$ can be written as,
\begin{equation}
\alpha = \frac{1}{\pi^2}\int_{0}^{\infty}dx \frac{\sqrt{x + \sqrt{2 x}}}{e^{\tilde{\beta}_{c}x} -1}.
\end{equation}
In the dilute regime $\alpha \ll 1$, a dip in the critical temperature at $\eta =1$ is observed due to the change in the energy dispersion. This drop in the critical temperature can be quantified by the ratio between the critical temperature at $\eta =1$ and that at $\eta=0$ (which is $T_{c} (\eta=0) = T^{0}_{c}/2^{1/3}$). Considering only the lowest brnach of dispersion, we obtain analytically the variation of relative drop in the critical temperature with dimensionless density,
\begin{equation}
T_{c} (\eta =1)/\epsilon_{0} = \tilde{T}_{1}\left[1 - \frac{\zeta(7/4)\sqrt{\tilde{T}_{1}}}{2 \sqrt{2}\zeta(5/4)}\right],
\label{tempeta1}
\end{equation}
where $\tilde{T}_{1} = \left[\frac{\pi^2 2^{1/4} \alpha}{\Gamma(5/4) \zeta(5/4)}\right]^{4/5}$.
\begin{figure}
\rotatebox{0}{\includegraphics*[width=8cm]{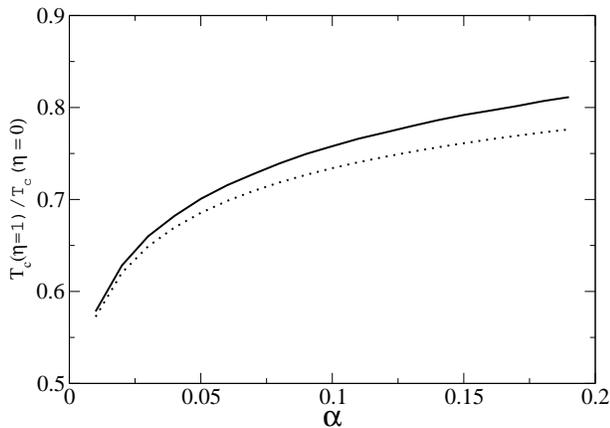}}
\caption{Relative drop of critical temperature $T_{c}$ at $\eta =1$ as a function of dimensionless density $\alpha$ for homogeneous gas of bosons with SO coupling. The dotted line represents the approximate analytical expression given in Eq.\ \ref{tempeta1}}
\label{fig3}
\end{figure}
In Fig.\ \ref{fig3} we compare the analytical estimate of the drop of $T_{c}$ with the result obtained numerically.
 
Interplay between the mean inter particle separation $\sim 1/\rho^{1/3}$ and the wavelength corresponding to the spin orbit coupling $\sim 1/k_{L}$ controls the behavior of the critical temperature as a function of $\eta$. 
As discussed above, we notice that in the dilute regime the SO interaction plays an important role and gives rise to a dip in the critical temperature at $\eta=1$. In the other limit, when the dimensionless density becomes much larger, the effect of SO interaction becomes less important. This can be understood from rewriting the Hamiltonian by scaling the lengths by mean interparticle separation,
\begin{equation}
H = \frac{2 \hbar^2 n^{2/3}}{m}\left[\frac{\vec{k}_{1}^2}{2m} + \frac{1}{\alpha^{1/3}}{k}_{1x} \sigma_{z} + \frac{\eta}{\alpha^{2/3}}\sigma_{x} \right]
\label{scale_ham}
\end{equation}
where $k_{1} = k/\rho^{1/3}$. For $\alpha \gg 1$, and $\eta > \alpha^{1/3}$, we can neglect the second term of the Hamiltonian and the critical temperature can be obtain from the expression,
\begin{equation}
\frac{T_{c}}{T^{0}_{c}} = \left[1 + \frac{g_{3/2}(z)}{\zeta(3/2)} \right]^{-2/3}.
\label{temp_hden}
\end{equation}
where $z = e^{-\eta \zeta(3/2)^{2/3}\beta_{c}/(\pi \beta^{0}_{c} \alpha^{2/3})}$, $g_{3/2}(z)$ is the usual function corresponding to Bose-Einstein integral and $\zeta$ is Riemann zeta function.
From this expression of $T_{c}$, we notice that for bose gas with higher density (high $\alpha$), $T_{c}$ approaches to its asymptotic value $T^{0}_{c}$ more slowly as seen in Fig.\ \ref{fig2}.

\section{Critical temperature of spin-orbit coupled Bose gas in a harmonic trap}
\begin{figure}
\rotatebox{0}{\includegraphics*[width=8cm]{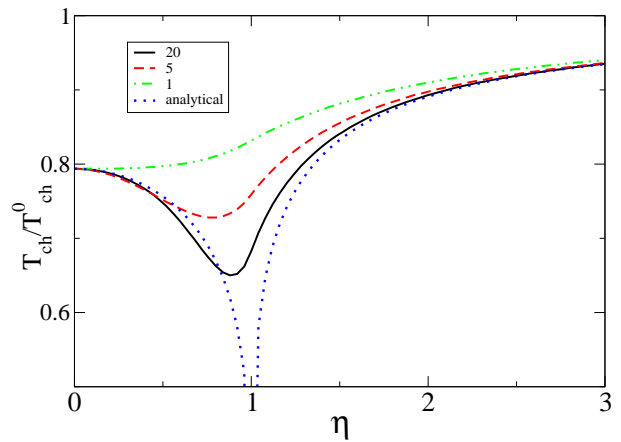}}
\caption{Variation of the scaled critical temperature $T_{c}/T^{0}_{c}$ with increasing Raman coupling $\Omega$ (in units of $\frac{\hbar^2 k_{L}^2}{m}$), for a system of harmonically trapped bose gas with SO coupling. Different lines represent the critical temperature for various values of dimensionless parameter $\chi$ related to the trap frequency and diluteness of the gas (values of $\chi$ are indicated in the graph). Analytical expressions obtained from Eq.\ \ref{crittempap} is represented by the dotted line. }
\label{fig4}
\end{figure}

For large number of bosons in a trap, the critical temperature for condensation $T_{ch}$ can be evaluated semiclassically,
\begin{equation}
N  =  \sum_{s} \int \frac{d^3p d^3r}{(2 \pi \hbar)^{3}} \frac{1}{ze^{\beta_{ch} E_{sc}(\vec{p},\vec{r},s)} -1},
\label{semiclassical1}
\end{equation}
where, $N$ is the number of bosons, $\beta_{ch} = \frac{1}{k_{B}T_{ch}}$, and
the fugacity can be calculated from ground state energy $z= e^{-\beta_{ch} E_{min}}$.For a slowly varying trap potential $V(\vec{r})$ the semiclassical energy is given by,
\begin{equation}
E_{sc}(\vec{p},\vec{r},s) = \frac{\vec{p}^{2}}{2m} + V(\vec{r}) + s\sqrt{\left(\frac{\hbar p_{x} k_{L}}{m}\right)^2 + \hbar^2 \Omega^2}.
\label{semiclass_en}
\end{equation}
Using Eq.\ \ref{semiclassical1}, and Eq.\ \ref{semiclass_en}, the critical
temperature $T^{0}_{ch}$ of harmonically trapped bosons without SO-interaction has been obtained analytically\cite{stringari_book} and is given by $k_{B} T^{0}_{ch} = \hbar \omega \left(\frac{N}{\zeta(3)}\right)^{1/3}$,
with $\omega$ is the frequency of the isotropic harmonic oscillator potential.
For SO interaction, the critical temperature can be obtained from the relation,
\begin{equation}
\zeta(3) = \sqrt{\frac{\chi}{2 \pi}}(\frac{T_{ch}}{T^{0}_{ch}})^{5/2}\sum_{s} \int_{-\infty}^{\infty}Li_{5/2}(e^{-\chi \frac{\beta_{ch}}{\beta^{0}_{ch}}\epsilon(\tilde{k}_{x},s)}) d\tilde{k}_{x},
\label{critical_ho}
\end{equation}
where $\chi = k_{L}^{2} l^{2} (\zeta(3)/N)^{1/3}$, harmonic oscillator length scale $l = \sqrt{\frac{\hbar}{m\omega}}$ and $Li$ is polylogarithmic function. The scaled critical temperature of SO coupled bec in a harmonic trap becomes a function of two dimensionless parameters $\chi$ , $\eta$, and can be written as 
$T_{ch}/T^{0}_{ch} = f(\chi, \eta)$. For $\eta =0$ this function f takes the value $1/2^{1/3}$ due to the spin degeneracy, and for $\eta \gg 1$, the value of the function $f \rightarrow 1$ asymptotically, since the spin is frozen along the x-axis due to the strong Raman coupling.
Within the effective mass approximation the critical temperature can be written as,
\begin{eqnarray}
T_{ch} =  (1 - \eta^2)^{1/6}T_{ch}^{0}/2^{2/3}~~ \text{for} \eta <1 \nonumber\\
T_{c} = (\frac{\eta -1}{\eta})^{1/6} T_{ch}^{0}~~ \text{for} \eta >1.
\label{crittempap}
\end{eqnarray}
As discussed in the previous section this approximation breaks down near $\eta \approx 1$ (see fig.\ \ref{fig4}). 

In a harmonic trap with finite number of bosons $N$, the parameter $N^{1/3}/k_{L}l$ becomes similar to the dimensionless density $\alpha$ in homogeneous gas. Another dimensionless parameter $k_{L}l$ represents the ratio between the oscillator length to the length scale introduced by SO coupling. For low trap frequency and for dilute system when $\chi \gg 1$, the cusp in critical temperature is smoothened out and minimum appears at $\eta \approx 1$ similar to what observed for homogeneous system of  bosons. With decreasing the value of $\chi$, the magnitude of relative drop in critical temperature reduces and also it shifts to smaller value of $\eta$ as depicted in Fig.\ \ref{fig4}. Below a certain value of the parameter $\chi \approx 1$, the critical temperature smoothly increases with $\eta$ and the dip in $T_{c}$ disappears. 

As discussed in the previous section that this particular feature of the critical temperature is due to the change in DOS at $\eta =1$ which is also related to the lifting of ground state degeneracy and change in shape of the energy dispersion.  
Unlike the homogeneous system, in a trap, the lifting of degeneracy of the ground states is a crossover phenomenon and does not occur exactly at $\eta =1$. The dimensionless trap parameter $\gamma = k_{L}l$ plays a crucial role in lifting of the ground state degeneracy.

To study the ground state degeneracy we numerically diagonalize the one dimensional Hamiltonian,
\begin{equation}
H_{1d} =  \frac{\hat{p}_{x}^2}{2m} + \frac{\hbar k_{L}}{m}\hat{p}_{x} \sigma_{z} + \frac{1}{2}m \omega^{2}x^{2} + \hbar \Omega \sigma_{x}
\label{ham_1d}
\end{equation}
From the eigenvalues we compute the energy gap $\Delta = (E_{2} - E_{1})/\hbar \omega $, between the ground state energy $E_{1}$ and the next excited state $E_{2}$. The variation of the energy gap with increasing $\eta$ for different trap frequencies are depicted in Fig.5a. 
For weak trap, when $\gamma \gg 1$ we recover almost the free particle result. In this case the ground state energy gap remains exponentially small and then increases smoothly above $\eta \approx 1$. Increasing the trap frequency (or decreasing the value of the parameter $\gamma$) shifts the crossover to lower value of $\eta$. Finally, below a critical value of the parameter $\gamma \approx 1$, the quasi degeneracy of the ground state vanishes and the energy gap monotonically increases (see Fig.5a). This phenomena is similar to the disappearance of minimum in the critical temperature for larger values of $\chi$. 
Also it is interesting to note that unlike the uniform system, the ground state degeneracy is always lifted up due to the presence of a trap below the crossover point. For weak trapping potential, the exponentially small splitting between the quasi-degenerate ground states can be understood from the quantum tunneling of a particle between two degenerate ground states. In a trap, the wavefunctions of the degenerate states overlap and the symmetric and antisymmetric combination of the wavefunctions represent the ground state and next excited state. The energy difference between these two states gives rise to the exponentially small energy splitting between the quasi-degenerate states. To estimate this energy gap in a large harmonic trap, we consider the variational ansatz,
\begin{equation}
\psi_{\pm} = {\cal{N}}\left[\left(\begin{array}{c} \cos \theta \\ -\sin \theta \\ \end{array}\right) e^{-ik_{1}x} \pm \left(\begin{array}{c} \sin \theta \\ -\cos \theta \\ \end{array}\right) e^{i k_{1} x}\right] e^{-x^{2}/2l^{2}},
\label{wavef_var}
\end{equation}
where, ${\cal{N}}$ is the normalization constant, $k_{1}$ and $\theta$ are the variational parameters which can be obtained by minimizing the energy of $\psi_{+}$ state. The expression for energy splitting is given by,
\begin{equation}
\frac{\Delta E}{\hbar \omega} = \frac{e^{-k_{1}^{2} l^{2}} \left[2 k_{1}^{2} l^{2} \sin 2\theta + 2 \eta \gamma^{2} \cos^{2}2\theta - \gamma  k_{1}l \sin4\theta\right]}{1 - \sin^{2}2\theta e^{-2 k_{1}^{2} l^{2}}}.
\label{var_gap}
\end{equation}
This gives a good estimate of the energy splitting between the quasi-degenerate ground states and the critical value of $\eta$ where the crossover takes place (as shown in Fig.5b). The energy gap obtained from the variational calculation increases very sharply at a critical value of $\eta$ indicating the crossover and the ansatz looses its meaning above the critical $\eta$.  
\begin{figure}
\centering
\includegraphics[width=8cm]{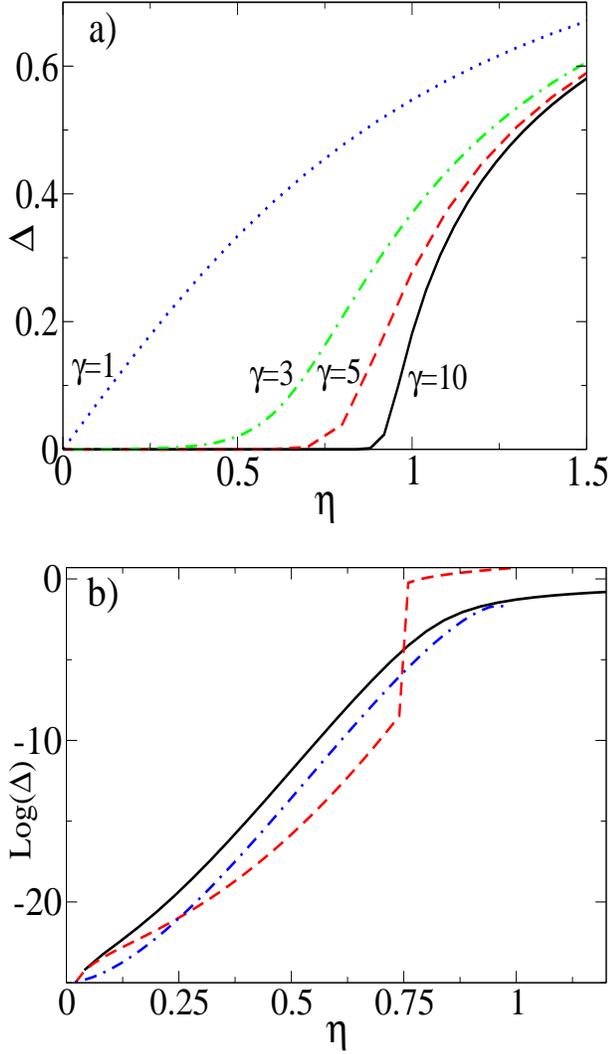}
\caption{a) Energy gap $\Delta$ above the ground state as the parameter $\eta$ increases by tuning the Raman frequency. Different curves are shown for decreasing values of the trap parameter $\gamma = k_{L}l$. b)Variation of logarithm of energy gap with $\eta$ for trap parameter $\gamma = 5$. Solid (black) line represents the exact result obtained from numerical diagonalization, energy gap obtained from the variational calculation is denoted by the dashed (red) line and the same calculated using WKB method (Eq.(\ref{wkb_gap})) is shown by the dot dashed (blue) line.}
\label{fig5}
\end{figure}

Analytical estimate of the energy gap between the quasi-degenerate ground states can be calculated by considering an effective one dimensional Hamiltonian in momentum space. For a smoothly varying trap, within the local density approximation the lower branch of the dispersion can be written as $E_{-} = \frac{p_{x}^{2}}{2m} + \frac{1}{2}m\omega^{2}x^{2} -\sqrt{\left(\frac{\hbar p_{x} k_{L}}{m}\right)^2 + \hbar^2 \Omega^2}$. Using the commutator $[\hat{p}_{x},x]=-i\hbar$, we can write $x= i\hbar \frac{\partial}{\partial p_{x}}$ and the lower branch of energies can be described by an effective Hamiltonian with an effective Planck constant,
\begin{equation}
H_{eff} = \epsilon_{0}\left[-\frac{\hbar_{eff}^{2}}{2}\frac{\partial^{2}}{\partial \tilde{p}_{x}^{2}} + V_{1d}(\tilde{p}_{x})\right],
\label{effective_pot}
\end{equation}
where, $\epsilon_{0} = \frac{\hbar^{2} k_{L}^2}{m}$,$\tilde{p}_{x} = p_{x}/\hbar k_{L}$, $\hbar_{eff}= 1/\gamma^{2}$ is the effective Planck constant and the one dimensional potential in momentum space is given by $V_{1d} = \frac{\tilde{p}_{x}^{2}}{2}  -\sqrt{\tilde{p}_{x}^{2} + \eta^{2}}$. It is interesting to note that the trapping frequency plays the role of the effective Planck constant and hence it gives rise to the quantum tunneling between the degenerate ground states formed at the minima of the classical potential at $p_{x} = \pm \sqrt{1 - \eta^{2}}$. Within the WKB approximation\cite{landau}, the energy splitting due to quantum tunneling can be written as,
\begin{equation}
\Delta E = \hbar \omega \sqrt{1- \eta^{2}}e^{-\alpha^{2}\left[\sqrt{1 - \eta^{2}} - \eta^{2}ln\{(1 + \sqrt{1 - \eta^{2}})/\eta\}\right]}.
\label{wkb_gap}
\end{equation}
This simple picture of quantum tunneling provides a clear picture of degeneracy lifting in a trap and the analytical value of the energy gap agrees well with the numerical result as shown in Fig.5b.
Both the disappearance of the minimum in the critical temperature and the lifting of the `quasi-degeneracy' are related crossover phenomena and are controlled by the dimensionless trap parameter $\gamma$.

\section{Bosons with large spin in a harmonic trap}
\begin{figure}
\rotatebox{0}{\includegraphics*[width=8cm]{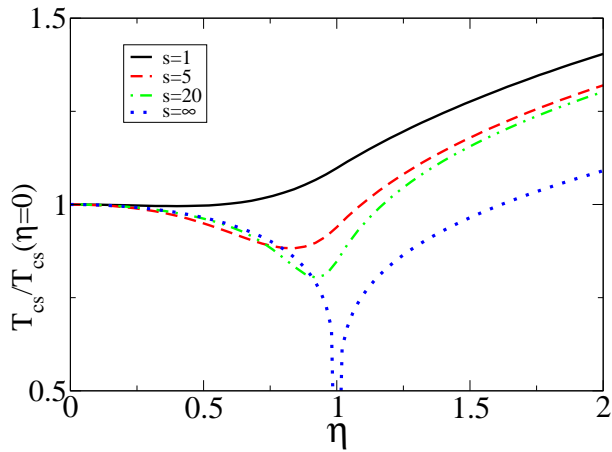}}
\caption{Variation of the scaled critical temperature of spin $s$ bosons $T_{cs}/T_{cs}(\eta =0)$ with increasing $\eta$, for fixed parameter $\chi = 3$. Different lines represent different spin $s$ of the bosons. Analytical result of the ctitical temperature $T_{\infty}/T_{\infty}(\eta=0)$ in the limit of $s \rightarrow \infty$  using the Dicke model (Eq.\ \ref{tcrit_inf}) is shown by the dotted line. }
\label{fig6}
\end{figure}

In this section we study the condensation of spin-orbit coupled bosons with large spin which is a natural extension of the recent experimental setup for atoms with two internal states. This is interesting for both theoretical modeling as well it provides a clear picture of the spin dependent phenomena observed in the experiment. 
The Hamiltonian describing harmonically trapped, spin-orbit coupled bosons with large spin of magnitude s can be written as,
\begin{equation}
H_{s} =  \frac{\hat{\vec{p}}_{\perp}^2}{2m} + \frac{1}{2}m \omega^{2}\vec{r}_{\perp}^{2} + H_{D},
\end{equation}
where $\vec{p}_{\perp}$, $\vec{r}_{\perp}$ are the momentum and position vector in the transverse directions (y,z coordinates). The spin-orbit interaction of a large spin $s$ is described by the Hamiltonian $H_{D}$,
\begin{equation}
H_{s} = \hbar \omega \left[a^{\dagger}a + i\frac{\gamma}{\sqrt{s}}(a^{\dagger} -a)S_{z} +2 \eta \gamma^2 S_{x}\right]
\label{dicke1}
\end{equation}
where $S_{z,x}$ are z, x components of the spin operator and $a$ is ladder operator of harmonic oscillator defined as $a= \sqrt{\frac{m\omega}{2 \hbar}}(\hat{x} + \frac{i}{m\omega \hbar}\hat{p}_{x})$.  Above Hamiltonian is similar to the Dicke model\cite{dicke} describing the `superradiant' quantum phase transition. For $s=1/2$, the Hamiltonian $H_{s}$ reduces to that in Eq.\ \ref{ham_1d} describing SO interaction of harmonically trapped pseudospin $1/2$ bosons. For large spin $s$, we can consider it as a classical vector with components $S_{x} = s\sin \theta \cos \phi$, $S_{y}= s\sin \theta \sin \phi$ and $S_{z} = s \cos \theta$. Within the semiclassical approximation the critical temperature $T_{cs}$ can be obtained from,
\begin{equation}
N = \frac{(2s +1)}{4 \pi}\int d\Omega \frac{d^3 p d^3 r}{(2 \pi \hbar)^3}  \frac{1}{e^{\beta_{cs}H_{s}(\vec{p},\vec{r},\theta,\phi)-E_{min}} - 1},
\label{crit_s1}
\end{equation}
where, $d\Omega = \sin\theta d \theta d \phi$ and $E_{min}$ is the energy of the classical ground state. In terms of the critical temperature of harmonic oscillator $T^{0}_{ch}$, the critical temperature of SO coupled bosons with spin s can be determined from the relation,
\begin{equation}
x^3 = \frac{1}{\zeta(3)}\frac{(2 s + 1)}{4 \pi}\int d\Omega g_{3}(e^{-\chi s x [e(\theta,\phi)-e_{min}]})
\label{selfcons_ts}
\end{equation}
where $x = T^{0}_{ch}/T_{cs}$, and the dimensionless parameter $\chi = \gamma^{2}(\zeta(3)/N)^{1/3}$ describes the diluteness of the gas as well as the strength of trapping frequency. From the classical energy for the spin configuration $e(\theta, \phi) = -\cos^2 \theta + 2 \eta \sin \theta \cos \phi$, the ground state energy is given by,
\begin{eqnarray}
e_{min} & = & - (1 + \eta^2),~\text{for}~ \eta <1,\nonumber\\
& = & - 2 \eta, ~\text{for}~ \eta > 1.
\label{min_es}
\end{eqnarray}  
For finite magnitude of spin, a minimum in the critical temperature appears near $\eta =1$. The relative drop in the critical temperature $T_{cs}$ is controlled by the dimensionless parameter $\chi$ and the magnitude of spin s. Below a certain value of $\chi s$ this minimum disappears and $T_{cs}$ increases monotonically.
 As shown in Fig.\ \ref{fig6}, for a fixed value of the parameter $\chi$ the relative drop in the critical temperature increases for larger values of spin s. Within the saddle point approximation the analytical expression of the critical temperature is given by,
\begin{eqnarray}
\frac{kT_{cs}}{\hbar\omega} & = &  \left[\sqrt{(1 - \eta^2)}\frac{2 s}{2 s + 1} \frac{N \gamma^2}{\zeta(4)}\right]^{1/4}, ~ \text{for}~ \eta < 1 \\ 
& = & \left[\sqrt{\eta(\eta -1)}\frac{4 s}{2 s + 1} \frac{N \gamma^2}{\zeta(4)}\right]^{1/4}, ~ \text{for}~ \eta > 1.
\label{tc_saddle}
\end{eqnarray}

In the limit of $s \rightarrow \infty$, the Hamiltonian $H_{D}$ similar to the Dicke model can be diagonalized by the Holstein-Primakof transformation\cite{brandes} and it can be written as,
\begin{equation}
H_{D} = \hbar \omega \left[\gamma^2 s e_{min} + \lambda_{-} C_{-}^{\dagger}C_{-} + \lambda_{+} C_{+}^{\dagger}C_{+} \right],
\label{diag_hamd}
\end{equation}
where $C_{-}$, $C_{+}$ are transformed ladder operators which couples the original operators of the harmonic oscillator with the spin excitations. Corresponding excitation frequencies $\lambda_{\pm}$ (in units of $\omega$) are given by,
\begin{eqnarray}
2 \lambda_{\pm}^2 & = &1 + 4 \eta^2 \gamma^4 \pm \sqrt{(1 - 4 \eta^2 \gamma^4)^2 + 16 \eta \gamma^4}~\text{for}~\eta >1 \nonumber\\
& = & 1 + 4 \gamma^4 \pm \sqrt{(1 - 4 \gamma^4)^2 + 16 \eta^2 \gamma^4}~\text{for}~\eta <1 
\label{fmodes}
\end{eqnarray}
The critical temperature $T_{\infty}$ of the SO coupled BEC in the limit of $s \rightarrow \infty$ can be determined from the four dimensional oscillator with excitation frequencies given in Eq.\ \ref{fmodes},
\begin{equation}
kT_{\infty} =  \hbar \omega  \left[\frac{N \lambda_{+} \lambda_{-}}{\zeta(4)}\right]^{1/4}.
\label{tcrit_inf}
\end{equation}
For $\eta > 1$, $T_{\infty}$ matches exactly with the critical temperature obtained from saddle point approximation (Eq.\ \ref{tc_saddle}) when $s \rightarrow \infty$. But for $\eta < 1$, $T_{\infty}$ is larger by a factor of $2^{1/4}$. This is because of two degenerate minima at $\cos \theta = \pm 
\sqrt{(1 - \eta^2)}$ for $\eta <1$ and both contributes in the saddle point approximation. Whereas for Dicke model with $s \rightarrow \infty$, the ground state chooses one of the minima and the frequencies $\lambda_{\pm}$ represents gaped excitation modes above the symmetry broken ground state. 
From this analysis it is clear that the dip in the critical temperature of SO coupled bose gas in a harmonic trap around $\eta \approx 1$ signifies a Dicke like quantum phase transition in the spin orientation. For bosons with finite spin the ground state is linear combination of degenerate minima and an energy gap always exists above the ground state which is controlled by the trap frequency and spin. Because of the energy gap, the critical temperature $T_{cs}$ of finite spin bosons does not vanish at $\eta =1$ and shows a smooth minimum. A true quantum phase transition occurs for infinite spin Dicke model and vanishing energy gap at $\eta =1$ leads to a sharp drop in the critical temperature for condensation to zero.

\section{Conclusion}
In summary, we investigated the effect of spin-orbit interaction on the transition temperature for Bose-Einstein condensation of non-interacting bosons. For uniform system of bosons the critical temperature at zero Raman coupling is reduced by a factor of $2^{2/3}$ due to two pseudospin degeneracy of the atoms. By increasing the Raman frequency $\Omega$, $T_{c}$ decreases upto a critical value $\Omega_c$ , then increases again and asymptotically reaches the value of the critical temperature of single component BEC.
At the critical value $\Omega_c$, the critical temperature drops to the minimum value, where the energy dispersion also changes its shape from double well structure to a single well. This cusp in $T_c$ at the critical Raman coupling $\Omega_c$ is an interesting feature of BEC with SO coupling which has already been observed in recent experiment\cite{ji}. 
This particular behavior of $T_{c}$ is due to the change in density of states and can be captured qualitatively by effective mass approximation.   The relative drop of $T_c$ at $\Omega_c$, depends on the dimensionless diluteness parameter $\alpha$ of the gas which is the ratio between the interparticle separation of the bosons and the wavelength associated with SO interaction. The $T_{c}$ at $\Omega_c$ decreases with decreasing the density of the gas. 

We also studied the critical temperature of harmonically trapped bosons with SO interaction within local density approximation. 
Sharp cusp in $T_{c}$ is smoothened out and the drop in $T_{c}$ exhibits a crossover phenomena which is controlled by the trap frequency. By increasing the diluteness of the gas or the trap frequency, the minimum of the critical temperature is reduced and shifted to the lower value of Raman frequency. Finally this cusp in $T_{c}$ disappears for sufficiently large trap frequency and $T_{c}$ increases monotonically with increasing $\Omega$. 
As observed in the uniform system, the kink structure in $T_{c}$ is also an demarcation between doubly degenerate ground states with equal and opposite momentum and a single ground state with zero momentum.  
Unlike the free system, the ground state is always non degenerate in presence of a trap for non vanishing $\Omega$. For low trapping frequency, the ground states are quasi-degenerate and separated by an exponentially small energy gap upto a critical $\Omega$ .
By increasing the Raman frequency, the exponentially small energy gap increases rapidly above a critical value, indicating a crossover from `quasi-degenerate' ground states to a single ground state. 
For increasing trap frequency this crossover occurs at smaller values of the Raman coupling and finally disappears above a critical trapping frequency ($\gamma \approx 1$), and the ground state becomes non degenerate for all values of $\Omega$. Both the minimum in $T_{c}$ and the lifting of ground state degeneracy shows similar crossover phenomenon which is controlled by the trapping frequency.
We also provide a clear picture and analytical estimate of the splitting between the quasi  degenerate ground states from quantum tunneling of a particle in an effective double well potential where the trapping frequency plays the role of the effective Planck constant.

Considering a `Dicke' like model, we extend our calculation for critical temperature to bosons with large spin. Similar to pseudospin $1/2$ bosons, a minimum in the critical temperature near $\Omega_c$ appears. Apart from the diluteness of the gas, spin of the bosons play an important role in this crossover phenomenon. At $\Omega_c$, $T_{c}$ decreases with increasing value of spin. Finally in the limit of infinite spin, the critical temperature vanishes as $|1 - \eta|^{1/8}$ at the critical point $\eta =1$ as a consequence of `Dicke' type quantum phase transition describing the spin ordering.
For bose gas with finite spin, this drop in $T_{c}$ is a manifestation of `Dicke' like quantum phase transition corresponding to the spin alignment, although the $T_{c}$ does not vanish at the critical point due to non vanishing energy gap. This phenomenon also signifies a change from `unconventional' condensate with quasi-degenerate ground states to a `conventional' one with single ground state.

\end{document}